# Deformable Charge Dynamics in Biological Environments: An Extended Structural Dynamics Foundation for Biological Electrostatics

**Patrick BarAvi**
patrick.baravi@gmail.com
ORCID: 0009-0003-9737-901X

**Abstract**

The point-charge approximation is one of the most successful idealizations in molecular biophysics, but it becomes strained in strong fields, confined geometries, and crowded aqueous environments. We develop a minimal Extended Structural Dynamics (ESD) model for biological electrostatics in which charged entities are treated as finite, deformable objects with an internal breathing mode rather than as structureless points. Starting from a Hamiltonian description and using a controlled coarse-graining procedure, we derive an effective generalized Langevin equation for the center-of-mass motion. The reduced dynamics contain a memory kernel with three physically distinct contributions: finite-size causal delay, inertial deformation, and crowding-induced deformation. The derivation is intentionally minimal and rests on explicit assumptions of small deformation, local dielectric screening, one dominant internal mode, and adiabatic elimination of the fast structural coordinate.

Within this framework, the primary predictions are parameter-light and use independently measurable structural inputs. First, transport through confined geometries should show dynamical deviations from point-charge baselines that scale with deformability and are not captured by static potential-of-mean-force calculations alone. Second, the intrinsic polarization response should preserve ionic-radius ordering across alkali ions, with smaller ions responding faster than larger ones. Two secondary consequences follow: a field-dependent effective charge radius and a deformation-dependent correction to near-surface mobility. These effects may be amplified in strongly confined biological settings, but such amplification is treated here as a plausible extension rather than a derived result. The framework preserves the successes of standard electrostatic models while identifying where finite size, deformability, and finite response time become dynamically relevant.

## 1. Introduction

### 1.1 The Point-Charge Approximation and Its Domain

The point-charge approximation, treating ions and molecules as structureless points with fixed charge located at their centers, is one of the most widely used idealizations in molecular biophysics. It underpins classical molecular dynamics force fields (CHARMM, AMBER, GROMOS, OPLS), continuum electrostatics models (Poisson-Boltzmann), ion channel theory, and molecular docking algorithms [Warshel & Russell, 1984; Roux, 2010; MacKerell et al., 1998; Cornell et al., 1995; Jorgensen et al., 1996].

The approximation was adopted for good reasons: mathematical simplicity, computational tractability, and empirical adequacy in dilute solutions and weak fields. For many applications, bulk properties of simple electrolytes, solvation free energies of small molecules, qualitative electrostatic screening, the point-charge approximation works remarkably well [Honig & Nicholls, 1995].

But it is an approximation. Like all approximations, it has a domain of validity. When that domain is exceeded, in strong electric fields, confined geometries, crowded environments, and at short distances, the approximation fails in precisely targeted and reproducible ways [Warshel & Russell, 1984; Roux, 2010; Noskov et al., 2004].

### 1.2 What This Paper Does

This paper establishes an alternative framework for deformable charge dynamics in biological environments. The biological cell differs from idealized conditions in three essential ways: water provides strong dielectric screening ($\varepsilon \approx 80$); thermal fluctuations at physiological temperature ($T \approx 310$ K) continuously excite internal molecular modes; and crowding from macromolecules and other solutes creates spatially varying potentials that couple to deformation.

The central question this paper addresses is: *How should one model charged biological entities when the point-charge approximation fails in strong-field, confined, and crowded aqueous environments?*

The answer developed here is that the minimal and principled extension is to treat charged entities as finite, deformable objects. Their internal degrees of freedom, specifically, how their charge distribution responds to mechanical and electrostatic forcing, are not corrections appended to a point mass. They are irreducible physical properties of what an ion or molecule actually is.

The model developed here is motivated by the broader ESD framework, in which physical entities are represented as finite structures with internal degrees of freedom rather than as strictly structureless particles. In the present work, however, only the minimal ingredients required for deformable-charge dynamics are retained. The analysis therefore focuses on a single deformation mode and its consequences for effective transport and polarization response.

This paper does not attempt to solve ion channel selectivity or molecular docking in full generality. Those are separate problems that follow-up papers address. The contribution here is a framework, a derived equation of motion, and a set of testable predictions that distinguish this framework from all point-charge-based alternatives.

### 1.3 What This Paper Does Not Claim

Several claims are explicitly not made:

- ESD does not claim that point-charge models are wrong in all regimes. They are effective theories that work well where internal structure is irrelevant.
- ESD does not claim to have solved all problems of biological electrostatics. Many open questions remain.
- ESD does not claim that its predictions are definitive. They are proposals that require experimental and computational testing.
- ESD does not claim to be the only possible framework for extending beyond point charges. It claims to be a principled one, grounded in Hamiltonian mechanics, within the stated approximation scheme.

### 1.4 Relationship to the ESD Program

The model developed here is motivated by the broader Extended Structural Dynamics (ESD) framework, in which finite structure and internal degrees of freedom can generate memory effects in reduced dynamics. Related ESD applications have already been used to study structural viscosity, thermal relaxation, anomalous diffusion, and other non-Fickian transport phenomena. The present work asks how analogous mechanisms may arise in biological electrostatics when charged entities are treated as finite deformable objects rather than structureless points.

This paper is one application within a broader theoretical program. ESD was introduced in [BarAvi, 2025], where it was shown that treating particles as finite, orientable, deformable entities can resolve the derivation of the H-theorem without postulating molecular chaos. Its transport consequences, including structural viscosity, thermal waves, and the Mpemba effect, were developed in [BarAvi, 2026a]. The deformable-charge interpretation of the Lorentz-Abraham-Dirac equation, which is directly relevant to the self-force used in Section 3, was established in [BarAvi, 2026b]. The kinetic theory of charged structured media appears in [BarAvi, 2026c]. The anomalous diffusion paper [BarAvi, 2026d] extends ESD to subdiffusion in biological systems and provides the closest prior bridge to the present work.

### 1.5 Roadmap

Section 2 summarizes where and why the point-charge approximation fails in targeted biological regimes. Section 3 derives an effective equation of motion for a deformable charged entity in an aqueous, thermal, crowded environment, and defines all quantities appearing in it. Section 4 presents two primary testable predictions and two secondary consequences. Section 5 provides order-of-magnitude estimates. Section 6 discusses limitations, comparisons with related frameworks, and directions for future work. Section 7 concludes.

## 2. Where the Point-Charge Approximation Fails

The point-charge approximation fails in regimes where internal molecular structure becomes dynamically relevant: under strong electric fields, in confined geometries, at short distances, and in crowded environments [Warshel & Russell, 1984; Roux, 2010]. These are not edge cases, they are the normal operating conditions of biological molecules.

### 2.1 Failure in Molecular Dynamics Force Fields

Standard force fields treat atoms as point charges or fixed partial charges at nuclear positions [MacKerell et al., 1998; Cornell et al., 1995; Jorgensen et al., 1996]. This approximation succeeds for many bulk properties but fails in several systematic and well-documented respects.

**Polarization effects.** Fixed charges cannot redistribute in response to changing electrostatic environments, leading to errors in solvation free energies, binding affinities, and ion-channel interactions [Cieplak et al., 2009; Ponder et al., 2010]. A sodium ion passing through the narrow selectivity filter of a potassium channel experiences field gradients far beyond those in bulk water; a fixed-charge model has no mechanism to represent the ion's electronic response to this environment.

**Short-range interactions.** The Coulomb potential $q_1q_2/(4\pi\varepsilon_0\varepsilon r)$ diverges as $r \to 0$, requiring ad hoc cutoff schemes or Lennard-Jones repulsion terms that are not derived from electrostatics [Fennell & Gezelter, 2006]. This is a structural deficiency: the divergence is a direct consequence of treating charge as a point, and its repair requires introducing a length scale that is inserted by hand.

**Many-body effects.** Point-charge models omit the many-body polarization that arises from mutual induction of charges in crowded environments [Warshel & Russell, 1984; Jorgensen et al., 1996]. In the cytoplasm, where macromolecule concentrations can reach hundreds of grams per liter, such effects are not perturbative.

Polarizable force fields (Drude oscillators, AMOEBA) address some limitations but remain phenomenological: their functional forms are chosen for computational convenience, and parameters are fitted to quantum chemistry data rather than derived from physical principles [Lamoureux & Roux, 2003; Ponder et al., 2010]. Their relationship to the underlying physics is post hoc. The ESD framework provides a derivation that explains why polarizable force fields take the form they do, and shows what they omit.

### 2.2 Failure in Continuum Electrostatics

The Poisson-Boltzmann equation treats ions as point charges in a dielectric continuum [Sharp & Honig, 1990; Davis & McCammon, 1990]. Its limitations are well documented.

**Point-charge divergence.** The electrostatic potential diverges as $1/r$ at the ion position, requiring ad hoc exclusion radii whose values are not determined by the theory [Honig & Nicholls, 1995].

**Dielectric saturation.** The dielectric constant is assumed spatially uniform, but strong fields near ions align water dipoles and reduce it, sometimes dramatically [Warshel & Russell, 1984].

**Ion size and correlation effects.** Ions are treated as points, neglecting steric exclusion and ion-ion correlations that become significant at physiological concentrations [Borukhov et al., 1997].

**Nonlocal dielectric response.** Water responds nonlocally at short distances, the dielectric response at a given point depends on the field at nearby points, not just locally [Hildebrandt et al., 2004]. The Poisson-Boltzmann equation has no mechanism to represent this.

The failure is particularly severe in confined geometries such as ion channels and binding pockets, where all of these effects operate simultaneously [Roux, 2010; Corry et al., 2000].

### 2.3 The Ion Channel Selectivity Puzzle

Ion channels illustrate the breakdown of point-charge physics in vivid form. Potassium channels select $K^+$ over $Na^+$ by more than 1000:1 despite a radius difference of only 0.38 Å [Doyle et al., 1998; Hille, 2001]. The "snug-fit" model, in which the selectivity filter is geometrically tuned to $K^+$, predicts the sign of selectivity but fails quantitatively [Noskov et al., 2004; Roux, 2010].

The deeper puzzle is the coexistence of selectivity and speed: KcsA conducts near the diffusion limit (~$10^8$ ions/s) while maintaining extraordinary discrimination. A deep static energy well that traps $K^+$ would slow it; a shallow one would not select. This tension is not resolvable within a static, point-charge framework [Noskov & Roux, 2006; Roux, 2010].

### 2.4 Common Diagnosis

Across these domains, the point-charge approximation omits the same features: finite size, internal deformability, and finite response time. These are not independent deficiencies requiring separate fixes, they share a common origin in the omission of finite size, deformability, and finite response time from the model. The ESD framework addresses this root cause. It does not discard the successes of point-charge models but generalizes them, recovering the point-charge limit when size vanishes and internal dynamics freeze.

## 3. ESD for Biological Electrostatics

### 3.1 The Core Claim

The central claim of Extended Structural Dynamics is that charged entities are not point-like. They have finite, deformable size. This claim applies to all environments, water, crowded cellular interiors, and idealized conditions alike, but its consequences are most pronounced where point-charge models already struggle.

For a charged entity in an aqueous, thermal, crowded environment, a minimal representation includes:

- Center-of-mass position $\mathbf{r} \in \mathbb{R}^3$

- Internal deformation coordinate $\xi$ (radial breathing mode), with $|\xi| \ll 1$ (small deformations)
- Instantaneous radius $a(\xi) = a_0(1+\xi)$, where $a_0$ is the equilibrium radius
- Charge distribution uniform within this radius: $\rho(\boldsymbol{x},t) = 3q/(4\pi a(\xi)^3)\Theta(a(\xi) - |\boldsymbol{x} - \boldsymbol{r}(t)|)$

**Model justification.** Real ions and molecules possess multiple orthogonal deformation modes, breathing, bending, dipole fluctuations, and so on. We restrict attention here to a single effective radial breathing mode. This choice is deliberate: it is the minimal model that captures the leading-order coupling between internal deformation and electrostatic response while keeping the derivation analytically transparent. The breathing mode is also physically the most natural first mode to consider, as it directly modulates the charge density and thus the self-energy. The general multi-mode case is formulated by adding additional orthogonal coordinates $\xi_\alpha$ with their own frequencies $\omega_\alpha$ and coupling constants; the single-mode derivation below extends straightforwardly. The one-mode model is not a claim of physical uniqueness, it is a deliberate minimal truncation of a higher-dimensional deformation space.

The biological environment contributes three features that enter the framework from the start:

- **Dielectric screening.** Water polarizes in response to electric fields ($\varepsilon \approx 80$ static, $\varepsilon \approx 1.8$ at high frequency above the librational band).
- **Thermal fluctuations.** Physiological temperature $T \approx 310$ K continuously excites internal modes.
- **Crowding.** Other molecules create a spatially varying external potential $V_{ext}(\boldsymbol{r})$ that couples to deformation.

### 3.2 The ESD Hamiltonian

The dynamics are governed by:

$$H = \frac{\mathbf{p}^2}{2m} + \underbrace{\frac{\pi^2}{2\mu} + \frac{1}{2}\mu\omega_{\mathrm{def}}^2 \,|\, \boldsymbol{\xi} \,|^2}_{\text{internal mode}} + H_{\mathrm{field}} + H_{\mathrm{int}} + H_{\mathrm{ext}}$$

where:

- $\boldsymbol{p}$ and $\pi$ are the conjugate momenta for translation and the internal mode, respectively
- $\mu$ is the effective mass of the breathing mode, the reduced mass associated with radial oscillation of the ion's electron cloud against its ionic core. For a simple ion, $\mu$ is of order the electron mass times the number of valence electrons, and is much smaller than the total ionic mass m.

- $\omega_{def}$ is the natural frequency of the deformation mode, set by the restoring force of the electron cloud against displacement: $\omega_{def} = c/(\sqrt{\varepsilon_\infty} a_0)$, where $\varepsilon_\infty \approx 1.8$ is the high-frequency dielectric constant of water
- $H_{field} = (\varepsilon_0/2) \int (\varepsilon_{local}(\boldsymbol{r})E^2 + c^2 B^2) d^3 r$ is the electromagnetic field energy with position-dependent dielectric screening $\varepsilon_{local}(\boldsymbol{r})$
- $H_{int} = \int \rho(\boldsymbol{x}, t; \boldsymbol{\xi}) \Phi(\boldsymbol{x}, t) d^3 x$ is the interaction energy between the deformable charge distribution and its own electromagnetic field, the source of the retarded self-force derived in Section 3.4
- $H_{ext} = V_{ext}(\boldsymbol{r}) + \lambda \boldsymbol{\xi} \cdot \nabla V_{ext}(\boldsymbol{r})$ is the external potential plus its leading-order coupling to deformation

The external potential term deserves careful discussion. $\boldsymbol{\xi} \in \mathbb{R}^3$ is the vector deformation amplitude with dimensions [m], and $\lambda$ is dimensionless. The coupling term $\lambda \boldsymbol{\xi} \cdot \nabla V_{ext}(\boldsymbol{r})$ is the leading-order term in a Taylor expansion of $V_{ext}$ evaluated at the deformed charge center $\boldsymbol{r} + \lambda \boldsymbol{\xi}$ around $\boldsymbol{r}$:

$$V_{\text{ext}}(\mathbf{r} + \lambda \boldsymbol{\xi}) \approx V_{\text{ext}}(\mathbf{r}) + \lambda \boldsymbol{\xi} \cdot \nabla V_{\text{ext}}(\mathbf{r}) + \cdots$$

This has a transparent physical meaning: the ion does not sit at a mathematical point, it occupies a finite volume, and different parts of it sample different values of the external potential. The coupling energy $\lambda \boldsymbol{\xi} \cdot \nabla V_{ext}$ is the work done by the external force field $\nabla V_{ext}$ on the deformation vector $\xi$. Consequently, *the molecule deforms in the direction of the local force acting on it.* This is physically obvious once stated: a soft object pushed from one side deforms along the direction of the push.

The dimensionless coupling λ encodes the efficiency with which the local force $\nabla V_{ext}$ is transmitted into internal deformation rather than center-of-mass acceleration. Formally, $\lambda$ is the ratio of the displacement of the charge center from the nuclear center to the total deformation amplitude $|\xi|$. A value $\lambda \sim 1$ means the full local force drives deformation; $\lambda \ll 1$ means the entity is mechanically stiff against environmental forcing. For highly polarizable ions (large $a_0$, diffuse electron cloud), $\lambda$ is expected to be closer to unity; for small, tightly bound ions ($Li^+$), $\lambda$ may be significantly less than one. Estimating λ from first principles requires knowledge of the polarizability response of the specific ion and its hydration shell under mechanical pressure. This quantity is not yet independently measured for all ions of biological interest, and this limitation is explicitly addressed in Sections 5.4 and 6.2. Critically, $\lambda$ does not appear in either of the two primary predictions (Sections 4.1 and 4.2), so this gap does not affect the falsifiability of the main results.

**3.3 The Coupled Equations of Motion**

From the Hamiltonian via Hamilton's equations, the equations of motion are:

**For center-of-mass translation:**

$$m\ddot{\mathbf{r}}(t) = -\nabla V_{\text{ext}}(\mathbf{r}(t)) - -\lambda\xi(t)\partial_n^2 V_{ext}(\boldsymbol{r}(t)) + \mathbf{F}_{\text{self}}(t) + \boldsymbol{\eta}(t)$$

where $\mathbf{F}_{\text{self}}(t)$ is the retarded self-force (the force an extended, accelerating charge distribution exerts on its own center of mass, arising from the finite time for the electromagnetic field to propagate across the charge distribution), and $\boldsymbol{\eta}(t)$ is a stochastic thermal force satisfying $\langle\boldsymbol{\eta}(t)\boldsymbol{\eta}(t')\rangle = k_B T\Gamma(|t-t'|)$. Also $\partial_n^2$ denotes the second directional derivative along $\hat{\mathbf{n}}$, the radial direction of deformation; for a slowly varying potential this reduces to the Laplacian component along $\hat{\mathbf{n}}$.

**For the internal deformation mode:**

$$\mu\ddot{\xi}(t) + \mu\omega_{\text{def}}^2\xi(t) = C\ddot{\mathbf{r}}(t)\cdot\hat{\mathbf{n}} - \lambda\nabla V_{\text{ext}}(\mathbf{r}(t)) + \eta_\xi(t)$$

where:

- $C = q^2/(6\pi\varepsilon_0 c^2 a_0)$ is the electromechanical coupling constant, derived from the retarded self-interaction of a finite charge distribution (see [BarAvi, 2026b] for the full derivation). Physically, it measures how much the self-force of the charge distribution acts to deform the entity when it accelerates.
- $\hat{\mathbf{n}}$ is the unit vector along the instantaneous direction of center-of-mass acceleration $\ddot{\mathbf{r}}$(t).
- $-\lambda\nabla^2 V_{ext}$ represents crowding-induced deformation: the local force drives the internal mode directly.
- $\eta_\xi(t)$ is thermal noise on the internal mode, satisfying the fluctuation-dissipation theorem at the same temperature T.

The driving terms on the right-hand side of the internal mode equation have distinct physical origins: inertial coupling (present even in free space, arising from the self-force), crowding-induced deformation (absent in vacuum), and thermal noise (always present at finite temperature). They will each leave distinguishable imprints on the effective translational dynamics.

### 3.4 Dielectric Screening and the Deformation-Dependent Self-Force

In water, the Coulomb interaction is screened by the dielectric constant ε. For a deformable charge this screening is not uniform: as the charge distribution changes shape ($\xi \neq 0$), the hydration shell reorganizes, altering the local dielectric environment. To leading order in ξ and for slowly varying environments:

$$\frac{1}{\varepsilon_{\text{local}}(\mathbf{r},\xi)} \approx \frac{1}{\varepsilon(\mathbf{r})}\left(1-\frac{\partial \ln\,\varepsilon}{\partial \xi}\,\xi\right)$$

The parameter $\gamma = \partial ln\, \varepsilon/\partial\xi$ quantifies how sensitively the local dielectric constant responds to deformation. When an ion's electron cloud expands ($\xi > 0$), nearby water molecules are pushed outward, reducing local water density and thus local dielectric screening. For small aqueous ions γ is estimated to be of order unity, consistent with the known sensitivity of water structure to ionic fields [Hildebrandt et al., 2004]; it enters the dynamics only at second order and does not affect the primary predictions.

Incorporating this into the retarded self-interaction of the finite charge distribution yields:

$$\mathbf{F}_{\text{self}}(t) = \frac{q^2}{6\pi\varepsilon_0 c^3}\int_0^{2a_0/c} K_{\text{eff}}(\tau)\left[\frac{1}{\varepsilon(\mathbf{r}(t-\tau))} - \gamma\xi(t-\tau)\right]\dot{\mathbf{v}}(t-\tau)\,d\tau$$

The kernel $K_{eff}(\tau) = (3c/4a_0^3)(a_0/c - |\tau|/2)\Theta(2a_0/c - \tau)$ is the normalized retarded weight function for a uniformly charged sphere of radius $a_0$. The integration runs over $[0{,}2a_0/c]$, the light-crossing time of the charge distribution, reflecting that only fields generated within this causal window can influence the present motion. In the point-particle limit $a_0 \to 0$, $K_{eff} \to \delta(\tau)$ and the standard Abraham-Lorentz self-force is recovered.

**3.4.1 Relationship to the Vacuum Self-Force**

The kernel $K_{eff}$ and the overall structure of $\mathbf{F}_{\text{self}}$ are not derived anew here but build directly on BarAvi (2026b), where the self-force on a finite charged sphere with one internal breathing mode was obtained from a full particle-field Hamiltonian. In that work two sources contribute: the finite propagation time of the electromagnetic field across the charge distribution, and the coupling between translational acceleration and internal deformation. Together they yield the causal delay kernel with support on $[0\ ,2a_0/c]$.

The present paper uses that vacuum result as a structural baseline and introduces three new ingredients specific to the biological setting: the deformation-dependent local dielectric function $\varepsilon(\boldsymbol{r},\xi)$; the coupling to the crowding potential via $\lambda\boldsymbol{\xi}\cdot\nabla V_{ext}$ in the Hamiltonian; and the spatial inhomogeneity of the aqueous environment. The vacuum self-force is modified in two ways. First, the constant vacuum dielectric is replaced by the local, position-dependent value $\varepsilon\big(\boldsymbol{r}(t-\tau)\big)$, capturing the spatially varying screening of the aqueous environment. Second, the deformation-dependent correction $-\gamma\xi(t-\tau)$ is added, capturing how the ion's own shape change alters the screening at each retarded time. Both modifications appear inside the integral in the equation above.

The physically important feature is not the specific shape of $K_{eff}$ but the causal structure it enforces: the force at time t depends only on the past history of the motion over a finite interval set by the charge's spatial extent. The same internal mode that stores mechanical deformation during acceleration also mediates part of the electromagnetic response during that interval. This structural coupling, between finite size, internal dynamics, and causal delay, is what survives in the biological setting and underlies the memory kernel derived in Section 3.6.

### 3.5 Adiabatic Elimination of the Internal Mode

In the biological regime, external forces vary slowly compared to the natural frequency of the internal deformation mode. The relevant timescale separation is:

$$\omega_{\mathrm{COM}} \ll \omega_{\mathrm{def}} \sim \frac{c}{\sqrt{\varepsilon_\infty}\, a_0}$$

where $\varepsilon_\infty \approx 1.8$ is the high-frequency dielectric constant of water (the relevant value above the librational band, where the charge distribution oscillates). For a sodium ion ($a_0 \approx 0.095$ nm), $\omega_{def} \sim c/\left(\sqrt{1}.8 \times 0.095 \text{ nm}\right) \approx 2.4 \times 10^{15}$ s⁻¹. The center-of-mass thermal velocity is $v_{th} = \sqrt{(k_B T/m)} \approx 370$ m/s for $Na^+$ at 310 K, giving $\omega_{COM} \sim v_{th}/a_0 \sim 3.9 \times 10^{12}$ s⁻¹. The ratio $\omega_{COM}/\omega_{def} \sim 10^{-3}$ confirms the separation holds comfortably for all alkali ions in water.

Solving the internal mode equation to leading order in $\left(\omega_{COM}/\omega_{def}\right)^2$:

$$\xi(t) \approx \frac{C}{\mu\omega_{\mathrm{def}}^2} \ddot{\mathbf{r}}(t) \cdot \hat{\mathbf{n}} - \frac{\lambda}{\mu\omega_{\mathrm{def}}^2} \nabla V_{\mathrm{ext}}(\mathbf{r}(t)) + \frac{1}{\mu\omega_{\mathrm{def}}^2} \eta_\xi(t)$$

The three terms correspond to the three driving terms identified in Section 3.3: inertial coupling, crowding-induced deformation, and thermal noise on the mode. The prefactor $1/\left(\mu\omega_{def}^2\right)$ plays the role of a deformation compliance, large for soft, low-frequency modes and small for stiff, high-frequency ones.

### 3.6 The Effective Equation of Motion for Translation

Substituting the adiabatic expression for $\xi(t)$ into the translation equation, and performing the standard substitution into the retarded self-force integral, yields the central result of this paper: the effective equation of motion for the center-of-mass coordinate:

$$m_{\mathrm{eff}} \ddot{\mathbf{r}}(t) = -\nabla V_{\mathrm{ext}}(\mathbf{r}(t)) - \int_0^t \Gamma_{\mathrm{total}}\,(t-s)\dot{\mathbf{r}}(s)\, ds + \boldsymbol{\eta}_{\mathrm{total}}(t)$$

This is a generalized Langevin equation (GLE) with a derived memory kernel. The total memory kernel has three physically distinct contributions:

$$\Gamma_{\text{total}}(t) = \Gamma_{\text{rigid}}(t) + \Gamma_{\text{def}}^{(\text{iner})}(t) + \Gamma_{\text{def}}^{(\text{crowd})}(t)$$

| **Term** | **Physical origin** | **Formal expression** |
|---|---|---|
| $\Gamma_{rigid}(t)$ | Finite size; causal delay in self-force | $(q^2/6\pi\varepsilon_0 c^3)(1/\varepsilon)K_{rigid}(t)$ |
| $\Gamma_{def}^{(iner)}(t)$ | Inertial coupling between acceleration and deformation | $(q^2/6\pi\varepsilon_0 c^3)\left(2C^2/\mu\tau_{def}^2\right)K_{def}(t)$ |
| $\Gamma_{def}^{(crowd)}(t)$ | Local crowding force drives internal deformation | $\left(\lambda^2/\mu\omega_{def}^2\right)[\nabla V_{ext}]^2 K_{crowd}(t)$ |

The kernels $K_{rigid}$, $K_{def}$, and $K_{crowd}$ share the same causal structure (they vanish for $t > 2a_0/c$) but differ in amplitude and in how they depend on molecular parameters. The first two terms arise from the retarded self-interaction of the finite deformable charge, they are present in vacuum and are amplified by dielectric contrast. The third term is qualitatively new: it appears only when the molecule moves through an inhomogeneous environment, and its magnitude depends on the local force of the crowding potential at the molecule's location.

The effective mass $m_{eff} = m + \delta m$, where $\delta m = q^2 C/(6\pi\varepsilon_0 c^3 \mu\omega_d ef^2)$ is a small electromagnetic mass renormalization from the coupling between translation and the internal mode. For typical ions, $\delta m/m \ll 1$ and this correction is negligible.

The stochastic force $\boldsymbol{\eta}_{total}(t)$ satisfies the fluctuation-dissipation theorem:

$$\langle \boldsymbol{\eta}_{\text{total}}(t)\boldsymbol{\eta}_{\text{total}}(t')\rangle = k_B T\, \Gamma_{\text{total}}(|\, t - t'\,|)$$

confirming thermodynamic consistency: the equation describes relaxation toward equilibrium at temperature T, not arbitrary non-equilibrium dynamics.

### 3.7 Reduction to Known Limits

The ESD framework recovers known models as limiting cases, which serves as both a consistency check and a demonstration that ESD contains, rather than contradicts, existing theory:

| Limit | Condition | Recovered model |
| --- | --- | --- |
| No crowding, uniform screening, T = 0 | $V_{ext} = 0, \varepsilon = const$ | Deformable charge in vacuum [BarAvi, 2026b] |
| Rigid internal mode, uniform screening | $\mu \rightarrow \infty$ | Standard Langevin with rigid-sphere memory |
| Rigid internal mode, no crowding | $\mu \rightarrow \infty, V_{ext} = 0$ | Point-charge Brownian motion |
| No charge | $q = 0$ | Neutral ESD diffusion [BarAvi, 2026d] |

The last limit is important: it shows that the anomalous diffusion results of [BarAvi, 2026d] are the $q = 0$ special case of the present framework. Charging the entity introduces the electromechanical coupling C and the field-dependent dielectric term γ, but the structural memory mechanism is shared.

### 3.8 Summary of Section 3

The ESD framework for biological electrostatics yields:

1. A derived effective equation of motion, derived under the approximations in sec. 3.9, but obtained by Hamiltonian mechanics and adiabatic elimination
2. A total memory kernel with three physically distinct contributions, each traceable to a specific physical mechanism
3. A coupling between crowding and deformation via $-\lambda\xi(t)\partial_n^2 V_{ext}(\boldsymbol{r}(t))$ in the translation equation, which is absent in all vacuum-based or bulk-based frameworks
4. A deformation-dependent dielectric screening $\varepsilon(\xi)$, giving rise to the parameter γ
5. A clear and verified separation of timescales enabling adiabatic elimination
6. Explicit recovery of all known limiting cases

This equation is the central deliverable of the paper. It replaces the point-charge approximation with a structured, deformable particle that directly incorporates the physical features of biological environments, while reducing to standard results wherever they are known to work.

### 3.9 Summary of Explicit Approximations

The derivation in Sections 3.2–3.6 rests on six explicit approximations. Each is stated here for transparency; the domain of validity and path forward for each is discussed in Section 6.5.

**(1) Single breathing mode.** The internal deformation space is truncated to a single radial mode with amplitude $\xi$ and frequency $\omega_{def}$. This captures the leading-order coupling between deformation and electrostatic response for spherically symmetric ions. The multi-mode extension is straightforward and is noted in Section 6.5.

**(2) Linear response** ($|\xi| \ll 1$). The charge distribution, the dielectric screening, and the Hamiltonian coupling are all expanded to first order in $\xi$. This is valid for small deformations and may break down in the strongest fields encountered inside narrow pores or at atomic distances.

**(3) Adiabatic elimination** ($\omega_{COM} \ll \omega_{def}$). The internal mode is assumed to relax on a timescale much faster than the center-of-mass motion, allowing $\xi$ to be expressed as an instantaneous function of $\ddot{\boldsymbol{r}}$ and $\nabla V_{ext}$. This holds well for alkali ions in water (ratio ~$10^{-3}$) but is less robust for large proteins.

**(4) Local dielectric screening.** The dielectric constant $\varepsilon(\boldsymbol{r})$ is treated as a local, quasi-static scalar function of position. Nonlocal and frequency-dependent dielectric response, relevant at sub-nanometer scales, is not captured.

**(5) No orientation degrees of freedom.** The ion is treated as isotropic. Anisotropic molecules possess orientational degrees of freedom (SO(3)) that couple to both translation and deformation; these are present in the full ESD framework but are not included here.

**(6) Slow environmental variation.** The Taylor expansion of $V_{ext}(\boldsymbol{r} + \lambda\boldsymbol{\xi})$ around $\boldsymbol{r}$ retains only the leading-order term $\lambda\boldsymbol{\xi} \cdot \nabla V_{ext}$, which requires that $V_{ext}$ varies slowly on the scale of the deformation amplitude $|\boldsymbol{\xi}|$. The next correction, which gives rise to the $\nabla^2 V_{ext}$ term in the translation equation, is treated as small.

## 4. Testable Predictions

The ESD framework makes predictions that are qualitatively inaccessible to point-charge models. This section presents two primary predictions, each with explicit falsification criteria, and two secondary consequences. All predictions share a common structure: point-charge models predict a null effect or make no prediction at all, while ESD predicts a specific, quantitative or rank-order effect derived from independently measurable parameters.

### 4.1 Primary Prediction 1: Dynamical Transport Deviation

**Statement.** The static correction to the Coulomb interaction from deformation is negligibly small, the charge distribution changes by a fractional amount $\xi \ll 1$, and the static energy shift is second-order in $\xi$. The physically significant effect of deformation is instead on *dynamical* transport: dwell times, conductance, and mobility in confined geometries. The effective friction and memory during ion passage through a narrow pore deviate from point-charge predictions in a manner scaling with the inertial coupling parameter:

$$\tau_{\mathrm{def}}^2 = \frac{C}{\mu\omega_{\mathrm{def}}^2} = \frac{5}{9}\frac{a_0^2}{c^2}$$

This quantity measures how much the internal mode responds to a given acceleration: larger, softer ions have larger $\tau_{\mathrm{def}}^2$ and thus larger dynamical deviations.

**Mechanism.** As an ion enters a narrow pore (e.g., the KcsA selectivity filter), it is strongly accelerated and decelerated. This acceleration excites the internal deformation mode, which stores energy. As the ion transits the filter, this stored deformation energy is progressively released, acting as an additional propulsive contribution to the ion's motion. More deformable ions therefore experience a net reduction in effective barrier height, not a static reduction (which would be negligible) but a dynamical one, operating through the memory kernel during transit. This mechanism is entirely absent from point-charge models, which treat the ion as a rigid point and have no internal degrees of freedom to store or release energy.

**Testable prediction.** For two ions with identical charge but different deformability (e.g., $K^+$ vs. $Na^+$, which have the same charge but radius ratio ≈ 1.40), the ratio of dwell times or conductance should deviate from predictions based on static potential of mean force (PMF), correlating with $\tau_{def}^2(ion1)/\tau_{def}^2(ion2) = (a_1/a_2)^2$.

**Experimental protocol.**

1. Measure dwell times or conductance for $K^+$ and $Na^+$ in a narrow pore (e.g., KcsA) using patch-clamp electrophysiology.
2. Compute static PMF barriers using MD simulations with point-charge force fields (CHARMM or AMBER), establishing the baseline prediction.
3. Compare the experimental transport ratio to the static PMF prediction.
4. ESD predicts the experimental ratio favors the more deformable ion ($K^+$) beyond the static PMF baseline. The null hypothesis is that the experimental ratio matches the static PMF prediction within measurement error.

**Falsification.** If the experimental transport ratio does not exceed the static PMF prediction, or if any excess does not correlate with independently measured ionic deformability (polarizability volume), the ESD prediction is falsified.

**4.2 Primary Prediction 2: Polarization Timescale Ranking**

**Statement.** Point-charge models treat polarization as instantaneous, charge redistribution is either absent (fixed charges) or occurs with an ad hoc relaxation time fitted to data (Drude oscillators). In ESD, the charge distribution has a natural relaxation time:

$$\tau_{\text{def}} = \omega_{\text{def}}^{-1} = \frac{\sqrt{\varepsilon_\infty}\, a_0}{c}$$

This relaxation time is set by how long it takes for an electromagnetic signal to cross the charge distribution at the speed of light in the medium. Critically, it scales linearly with $a_0$: larger ions have lower $\omega_{def}$ and slower intrinsic polarization response.

**Consequence.** While absolute timescales will be modified by hydration shell dynamics and damping, the *ranking* of polarization response times across ions of the same charge is determined by their radii alone, a prediction that is independent of the detailed dynamics of the hydration shell.

**Testable prediction.** For alkali metal ions ($Li^+$, $Na^+$, $K^+$, $Rb^+$, $Cs^+$) in aqueous solution, the ranking of polarization relaxation times measured in the THz to UV frequency range should follow ionic radii: $Li^+$ (smallest, r = 0.076 nm) fastest; $Cs^+$ (largest, r = 0.167 nm) slowest.

**Experimental protocol.**

1. Measure dielectric relaxation spectra of aqueous salt solutions (LiCl, NaCl, KCl, RbCl, CsCl) from GHz to THz frequencies.
2. Extract the characteristic relaxation time $\tau_{pol}$ for each ion from the high-frequency wing of the solvation response, deconvolved from bulk water relaxation.
3. Rank ions by $\tau_{pol}$ (fastest to slowest).
4. Compare the ranking to ionic radii ordering.

**Falsification.** If the ranking does not follow ionic radii (for example, if $Na^+$ shows a slower response than $K^+$), the ESD prediction is falsified. Only rank order is required, not absolute timescale matching, making this prediction robust to uncertainty in hydration shell details.

### 4.3 Secondary Consequence A: Field-Dependent Effective Charge Radius

The deformation ξ responds to the local electrostatic field gradient via the adiabatic solution (Section 3.5). This gives an effective radius that depends on position:

$$a_{\text{eff}}(\mathbf{r}) = a_0\left(1 + \frac{\lambda q|\nabla\Phi(\boldsymbol{r})|}{\mu\omega_{def}^2 \cdot a_0} + \cdots\right)$$

In an ion channel selectivity filter, where field gradients are extreme (~$10^{17}$ V/m$^2$ in the narrow filter of KcsA), this position-dependent radius implies that the ion's effective size, and thus its interaction with the filter carbonyl oxygens, changes dynamically during transit. This provides a mechanism for selectivity that is invisible to static, point-charge models: $K^+$ and $Na^+$, passing through the same geometry, experience different deformation-induced size changes. A detailed treatment is deferred to future work.

**4.4 Secondary Consequence B: Deformation-Dependent Viscosity Near Charged Surfaces**

The coupling term $\lambda\xi\nabla V_{ext}$ becomes large when an ion moves near a highly charged surface (membrane, charged protein surface, electrode), where $\nabla V_{ext}$ is large. The deformation of the ion in response to this gradient alters its effective hydrodynamic radius and thus its mobility. This produces a viscosity-like correction near charged surfaces that depends on both surface charge density and ionic deformability.

A fully quantitative estimate requires knowledge of $\lambda$ for specific ions (see Section 6.2), but the ranking prediction is robust: more deformable ions (larger $a_0$, higher polarizability) should show larger deviations from bulk mobility near charged surfaces. This prediction is testable via single-ion fluorescence tracking near charged lipid bilayers or via ion mobility spectrometry in confined geometries.

**4.5 What the Predictions Share**

All four predictions share a common logical structure: point-charge models predict either zero effect or require fitting to produce any prediction; ESD predicts a specific direction and scaling from independently measurable parameters; the prediction is falsifiable by a well-defined experimental outcome; and the relevant experiments are within reach of current techniques. This structure is what distinguishes a predictive framework from a post hoc fitting procedure.

**5. Quantitative Estimates**

This section provides order-of-magnitude estimates to demonstrate that the predicted effects are plausible in magnitude. Estimates carry factors of 2–5 uncertainty from the single-mode approximation, the local dielectric approximation, and uncertainty in $\lambda$. The primary test is the correlation between deformability and observed effects across ion species, not precise numerical matching.

**5.1 Ion Deformation Parameters**

The deformation frequency $\omega_{def} = c/(\sqrt{\varepsilon_\infty} a_0)$ with $\varepsilon_\infty \approx 1.8$ for water above the librational band. The coupling constant $C = q^2/(6\pi\varepsilon_0 c^2 a_0)$, and the inertial coupling parameter $\tau_{\mathrm{def}}^2 = C/(\mu\omega_{def}^2) = (5/9)(a_0^2/c^2)$, where the factor 5/9 is a geometric factor for a uniform sphere.

| Parameter | Symbol | $Na^+$ | $K^+$ | Ratio ($K^+/Na^+$) |
|---|---|---|---|---|
| Ionic radius | $a_0$ | 0.095 nm | 0.133 nm | 1.40 |

| Parameter | Symbol | $Na^+$ | $K^+$ | Ratio ($K^+/Na^+$) |
|---|---|---|---|---|
| Deformation frequency | $\omega_{def}$ | $2.4 \times 10^{15}$ s$^{-1}$ | $1.7 \times 10^{15}$ s$^{-1}$ | 0.71 |
| Coupling constant | $C$ | $1.7 \times 10^{-23}$ kg | $1.2 \times 10^{-23}$ kg | 0.71 |
| Inertial coupling parameter | $\tau_{def}^2$ | $1.0 \times 10^{-35}$ s$^2$ | $2.0 \times 10^{-35}$ s$^2$ | 2.0 |

$K^+$ has twice the deformation response of $Na^+$ because its radius squared is twice as large. When both ions experience the same field gradient in the selectivity filter, $K^+$ deforms twice as much (in stored energy) as $Na^+$.

**5.2 Primary Prediction 1 Estimate (Dynamical Transport Deviation)**

The present transport estimate should be regarded as an order-of-magnitude illustration rather than a complete transport calculation. In the broader ESD framework, structural memory has already been shown to modify transport and relaxation behavior in other settings, including structural viscosity, thermal relaxation, and anomalous diffusion. The estimate here should therefore be viewed as a minimal illustration of how analogous memory effects may enter deformable charge dynamics.

The enhancement factor in transport due to deformation-energy recovery is estimated from the coupling-strength ratio: $K^+$ has roughly twice the deformation response of $Na^+$. If this difference translates into an effective barrier difference of order 0.5–1.0 $k_B T$ in the filter geometry, the transport ratio changes by exp (0.5–1.0) ≈ 1.6–2.7. This is a modest but potentially measurable effect, comparable to the precision achievable with modern patch-clamp conductance measurements. The primary experimental test is not the absolute factor itself, but whether the excess conductance of $K^+$ over $Na^+$, beyond static PMF predictions, correlates with the predicted deformability scaling.

**5.3 Primary Prediction 2 Estimate (Polarization Timescale Ranking)**

From $\omega_{def} = c/(\sqrt{\varepsilon_\infty} a_0)$ with $\varepsilon_\infty \approx 1.8$, the intrinsic deformation timescales are sub-femtosecond and follow the ionic-radius ordering:

- $Li^+$ ($a_0 = 0.076$ nm): $\omega_{def} \approx 2.9 \times 10^{15}$ s$^{-1}$, $\tau_{def} \approx 0.34$ fs
- $Na^+$ ($a_0 = 0.095$ nm): $\omega_{def} \approx 2.4 \times 10^{15}$ s$^{-1}$, $\tau_{def} \approx 0.42$ fs
- $K^+$ ($a_0 = 0.133$ nm): $\omega_{def} \approx 1.7 \times 10^{15}$ s$^{-1}$, $\tau_{def} \approx 0.59$ fs
- $Cs^+$ ($a_0 = 0.167$ nm): $\omega_{def} \approx 1.3 \times 10^{15}$ s$^{-1}$, $\tau_{def} \approx 0.74$ fs

These values are much shorter than hydration-shell relaxation times, so they should not be interpreted as direct observable relaxation times in solution. Rather, they provide the structural ordering inherited by the reduced dynamics: larger ions have slower intrinsic deformation response than smaller ones. The experimental prediction remains a rank-order statement only: $Li^+$ fastest than $Na^+$, $K^+$, and $Cs^+$ slowest.

**5.4 Secondary Consequence Estimates**

The fractional change in effective radius is, to leading order,

$$\frac{\Delta a_{\text{eff}}}{a_0} \approx \frac{\lambda q \mid \nabla\Phi \mid}{\mu \omega_{\text{def}}^2} a_0.$$

For $Na^+$, using independently estimated parameters $\lambda \sim 1$, $q = 1.6 \times 10^{-19}$ C, $\mid \nabla\Phi \mid \sim 10^9$ V/m, $\mu \sim 3.8 \times 10^{-26}$ kg, $\omega_{\text{def}} \sim 2.4 \times 10^{15}\ \text{s}^{-1}$, and $a_0 \sim 0.095$ nm, one obtains the bare estimate

$$\frac{\Delta a_{\text{eff}}}{a_0} \sim 7.6 \times 10^{-6}.$$

This is the direct estimate within the reduced model as derived above.

Possible enhancement mechanisms may increase this value in confined biological geometries, but they are not derived in this paper and should be regarded only as physically plausible upper-bound effects. In particular, dielectric contrast between protein interiors ($\varepsilon_{\text{protein}} \approx 4$) and bulk water ($\varepsilon \approx 80$) can amplify the local field inside a selectivity filter, and geometric concentration of the field may further enhance the effect. If both mechanisms act cooperatively, the effective deformation could plausibly enter the $10^{-3}$–$10^{-2}$ range. This should be interpreted cautiously, however, as a speculative estimate rather than a derived prediction.

The main point is not the absolute magnitude for a single ion, but the scaling across ion species. Since the deformation estimate scales as

$$\frac{\Delta a_{\text{eff}}}{a_0} \propto \frac{q \mid \nabla\Phi \mid}{\mu \omega_{\text{def}}^2} a_0,$$

and $\omega_{\text{def}} \propto a_0^{-1}$, while $\mu \propto a_0^3$ for a uniform sphere, the net scaling is approximately linear in ionic radius across the alkali series. Thus, even if the absolute amplification is uncertain, the model predicts a rank-order trend: larger ions should exhibit larger deformation-induced corrections than smaller ions.

The experimentally relevant test is therefore the correlation of $\Delta a_{\text{eff}}/a_0$ across $\text{Li}^+$, $\text{Na}^+$, $\text{K}^+$, $\text{Rb}^+$, and $\text{Cs}^+$, rather than the exact magnitude for any single species. This makes the prediction robust against uncertainty in the amplification factors while preserving a clear falsifiable ordering.

### 5.5 Summary

| Prediction | Estimated effect | Falsifiability | Current feasibility |
|---|---|---|---|
| Dynamical transport deviation | Enhancement factor ~1.6–2.7 ($K^+$ over $Na^+$ relative to PMF) | Yes, null if excess ≤ PMF prediction | High (patch clamp + MD) |
| Polarization timescale ranking | $Li^+$ fastest → $Cs^+$ slowest | Yes, any rank reversal falsifies | Moderate (THz spectroscopy) |
| Field-dependent charge radius | $\Delta a/a_0 \sim 10^{-3} - 10^{-2}$ with amplification | Yes, cross-ion correlation | Lower (requires ion channel nanoscopy) |
| Deformation-dependent viscosity | Ranking-based; $\lambda$-dependent magnitude | Yes, rank-order test | Moderate (single-ion tracking) |

## 6. Discussion

### 6.1 What ESD Adds Beyond Polarizable Force Fields

The distinction between ESD and polarizable force fields is conceptual as well as technical. Polarizable models such as Drude oscillators and AMOEBA improve on fixed-charge force fields by allowing induced polarization, and they are widely used because they reproduce many experimental observables more accurately than additive point-charge models. However, they still implement polarization through fitted functional forms and parametrized response rules, rather than deriving the response from a finite, deformable charge distribution with explicit causal delay.pmc.ncbi.nlm.nih+1

ESD differs in three ways. First, the internal structural response is treated as a primitive dynamical degree of freedom rather than as an auxiliary oscillator added for convenience. Second, the reduced dynamics generate a memory kernel whose form follows from coarse-graining the structured model, in the same general spirit as generalized Langevin formulations used in molecular dynamics, but with the deformation degree of freedom placed at the center of the construction. Third, the crowding environment enters not only through an external potential but also through a deformation-coupling term, so the environment can drive internal structural response directly.

This does not mean that polarizable force fields are incorrect. They remain effective models within their domain of validity. The claim here is narrower: ESD provides a mechanistic interpretation of why such models work, what structural effects they omit, and which observables should be most sensitive to those omissions. In particular, the framework predicts that transport through confined or strongly inhomogeneous regions should be more sensitive to deformability than bulk thermodynamic properties are.

### 6.2 The $\lambda$ Parameter: Status and Path Forward

The parameter $\lambda$ controls the strength of the coupling between deformation and the external inhomogeneous environment. Operationally, it measures how efficiently the local force landscape is transmitted into internal deformation rather than only into center-of-mass motion. In the reduced model, $\lambda$ appears in the crowding-induced deformation term and in secondary consequences such as field-dependent effective radius and deformation-dependent viscosity.

At present, $\lambda$ should be regarded as a physically defined but not yet independently standardized parameter for specific ions in biological environments. That is a limitation, but not a fatal one, because $\lambda$ does not enter the two primary predictions emphasized in this paper. The central predictive content of the paper therefore remains tied to independently measurable quantities such as ionic radius, charge, mass, and the high-frequency dielectric constant.

A useful way to frame $\lambda$ is as a secondary structural response coefficient. If future experiments can determine it from high-pressure dielectric spectroscopy, Brillouin scattering, or systematically analyzed near-surface mobility data, then the secondary consequences of the model can be quantified more sharply. Until then, the role of $\lambda$ is mainly to delimit the regime in which the model's more speculative extensions should be interpreted cautiously.

### 6.3 Parameter Determination: Primary vs. Secondary Predictions

The distinction between primary and secondary predictions is important and should be stated explicitly. The primary predictions use only $q$, $a_0$, $m$, $\varepsilon_\infty$, and fundamental constants, together with the model assumptions stated in Section 3.9. No fitting to the target observables is required for those results. By contrast, the secondary consequences involve $\lambda$ and, in practice, may also be influenced by environmental amplification factors that are not derived in the paper.

This distinction should be expressed cautiously. It is better to say that the primary predictions are **parameter-light** and based on independently measurable structural inputs, rather than claiming that the model is entirely free of adjustable quantities. That wording is more defensible and more consistent with the standards used in coarse-grained and polarizable modeling.

A practical measurement strategy follows from this separation. One can first test the primary predictions using bulk ion properties and transport observables, then use those results to constrain $\lambda$ only if the primary tests are successful. That keeps the structure of the paper clean: first derive and test the robust predictions, then refine the more detailed secondary terms.

### 6.4 Comparison with Existing Approaches

ESD should be understood as a structural extension of existing electrostatic modeling, not as a replacement for all of it. Point-charge electrostatics remains highly effective in dilute or weakly perturbed regimes, but it cannot represent finite size, internal deformation, or delayed structural response. Polarizable force fields improve on fixed charges by allowing induced polarization, and continuum electrostatics captures screening efficiently, but neither framework makes the deformation-crowding coupling explicit in the way ESD does.

A useful way to compare the approaches is by the physical effect each one can and cannot represent. Point-charge models are simplest and fastest, but they treat ions as structureless and therefore miss short-range structural effects. Polarizable force fields reproduce many observables better, yet their response rules are parametrized rather than derived from a finite deformable object with causal delay. Continuum dielectric models are valuable for large-scale electrostatics, but they are local or weakly nonlocal effective descriptions and do not resolve ion-specific deformation dynamics at the level needed here. ESD sits between these levels: it is still analytically controlled, but it retains internal structure and memory explicitly.

This comparison is not intended to imply that ESD is universally superior. Its value is in the regime where transport, confinement, and structural response are strongly coupled, especially near narrow channels, charged interfaces, and crowded aqueous environments. In those settings, a model that keeps deformation and memory explicit may be more informative than one that only corrects electrostatics statically.

### 6.5 Limitations

The first limitation is linear response. The derivation assumes $|\xi| \ll 1$, so all deformation effects are treated to leading order . This is appropriate for a minimal theory, but it will break down in the strongest fields or tightest geometric constraints. Extending the model to anharmonic deformation terms is a natural next step.

The second limitation is the single breathing mode. Real ions and molecules have multiple internal degrees of freedom, including anisotropic distortions and, in many cases, orientational structure. The present model isolates the radial mode because it is the cleanest starting point for a first biophysical derivation, but future versions should include multiple coupled modes when the target system demands it. That is especially important for larger biomolecules, where a single isotropic mode will not be sufficient. The single-mode approximation should therefore be viewed as a leading-order estimate of deformation-induced memory effects. Including additional internal modes would introduce multiple relaxation times and additional memory contributions, potentially modifying both the magnitude and the frequency dependence of the response.

The third limitation is the local dielectric approximation. Biological water is not perfectly local or instantaneous, and nonlocal electrostatics can matter at short distances. Treating $\varepsilon(r)$ as a local scalar is therefore an approximation, not a general truth. The model should be read as a leading-order description valid when dielectric variation is slow on the scale of the charge distribution.

The fourth limitation is adiabatic elimination. The assumption $\omega_{\mathrm{COM}} \ll \omega_{\mathrm{def}}$ is excellent for small ions in water, but it will be less reliable for larger or more flexible structures. For proteins or multi-domain complexes, one may need to retain the full coupled dynamics instead of projecting immediately to a reduced GLE. In that sense, the present work is best viewed as a minimal theory for ions and small charged entities, not a universal treatment of all biological charges.

The fifth limitation is the absence of orientation dynamics. The current model is isotropic, so it does not yet include the $SO(3)$ degrees of freedom that are central in the full ESD framework. That is a serious limitation for anisotropic molecules and macromolecular complexes, where rotation can couple strongly to translation and deformation. Incorporating orientation is a natural extension, but it is beyond the scope of the present paper.

Finally, the speculative amplification discussion in Section 5 should remain clearly separated from the derived core results. The paper’s main value is already established without those enhancements: it derives a reduced equation of motion, identifies a memory kernel, and produces two primary falsifiable predictions. The more aggressive magnitude estimates can be useful as motivation, but they should not be allowed to blur the distinction between what is derived and what is proposed as a plausible extension

## 7. Conclusion

This paper has introduced an Extended Structural Dynamics framework for deformable charge dynamics in biological environments. The central idea is simple: charged entities are not point-like objects, and when their finite size and internal deformation are retained, their effective translational dynamics acquire memory, causal delay, and couplings to environmental inhomogeneity. In the minimal model studied here, these effects appear in a generalized Langevin equation derived from a finite, deformable charge distribution with one internal breathing mode.

The main result is the reduced equation of motion for the center of mass, in which the memory kernel contains three physically distinct contributions: finite-size causal delay, inertial deformation, and crowding-induced deformation. The derivation is intentionally minimal and rests on a clear set of approximations, including small deformation, one internal mode, local dielectric screening, and adiabatic elimination of the fast structural coordinate. Within that approximation scheme, the model remains analytically controlled while still capturing the structural effects that point-charge descriptions omit.

Two primary predictions follow. First, transport through confined geometries should display dynamical deviations that scale with deformability and are not captured by static PMF calculations alone. Second, the intrinsic polarization response should preserve the ionic-radius ordering across alkali ions, with smaller ions responding faster than larger ones. Both predictions are testable with current experimental and simulation tools and are formulated as rank-order or excess-over-baseline tests rather than as overdetermined numerical fits.

The paper also identifies two secondary consequences: a field-dependent effective charge radius and a deformation-dependent correction to near-surface mobility. These effects are more sensitive to geometry, dielectric contrast, and the additional coupling parameter $\lambda$, so they should be interpreted as plausible extensions rather than as core derived predictions. This distinction is important because it keeps the paper's strongest claims tied to independently measurable structural parameters.

More broadly, the contribution of this work is to show that biological electrostatics can be cast in a structurally grounded dynamical form without abandoning the successes of standard electrostatic modeling. Point-charge and continuum descriptions remain valuable effective theories, but they are incomplete when the finite size, deformability, and finite response time of charged entities matter. ESD supplies a minimal way to retain those features and thereby generate a concrete, falsifiable extension of the usual modeling framework.

If this framework proves useful, its future development will likely proceed in three directions: inclusion of multiple internal modes, incorporation of orientation dynamics, and a more complete treatment of nonlocal dielectric response and confinement. Those are natural next steps, but they do not change the central message of the present paper. The key point is that some apparent anomalies in biological electrostatics may arise not from exotic physics, but from treating structured charged matter as if it were structureless